\documentclass[12pt,a4paper,english,superscriptaddress]{revtex4}
\usepackage{graphics}
\usepackage{amsmath}
\usepackage{dcolumn}
\usepackage{amssymb}
\usepackage{bm}
\usepackage{graphicx}
\makeatletter
\usepackage{babel}
\usepackage[latin1]{inputenc}
\usepackage{hyperref}
\renewcommand{\a}{\alpha}
\renewcommand{\b}{\beta}

\usepackage{color}

\def\m{\mu}

\def\r{\rho}

\def\t{\tau}

\def\x{\xi}

\def\L{\Lambda}

\def\As{A\!\!\!/}
\def\bs{b\!\!\!/}
\def\ps{p\!\!\!/}
\def\ks{k\!\!\!/}
\def\as{a\!\!\!/}

\def\ds{\partial\!\!\!/}

\def\n{\nu}
\def\m{\mu}
\def\n{\nu}

\newcommand{\be}{\begin{equation}}
\newcommand{\ee}{\end{equation}}
\newcommand{\bea}{\begin{eqnarray}}
\newcommand{\eea}{\end{eqnarray}}

\newcommand{\pa}{\partial}

\begin{document}

\immediate\write16{<WARNING: FEYNMAN macros work only with emTeX-dvivers
                    (dviscr.exe, dvihplj.exe, dvidot.exe, etc.) >}
\newdimen\Lengthunit
\newcount\Nhalfperiods
\Lengthunit = 1.5cm
\Nhalfperiods = 9
\catcode`\*=11
\newdimen\L*   \newdimen\d*   \newdimen\d**
\newdimen\dm*  \newdimen\dd*  \newdimen\dt*
\newdimen\a*   \newdimen\b*   \newdimen\c*
\newdimen\a**  \newdimen\b**
\newdimen\xL*  \newdimen\yL*
\newcount\k*   \newcount\l*   \newcount\m*
\newcount\n*   \newcount\dn*  \newcount\r*
\newcount\N*   \newcount\*one \newcount\*two  \*one=1 \*two=2
\newcount\*ths \*ths=1000
\def\GRAPH(hsize=#1)#2{\hbox to #1\Lengthunit{#2\hss}}
\def\Linewidth#1{\special{em:linewidth #1}}
\Linewidth{.4pt}
\def\sm*{\special{em:moveto}}
\def\sl*{\special{em:lineto}}
\newbox\spm*   \newbox\spl*
\setbox\spm*\hbox{\sm*}
\setbox\spl*\hbox{\sl*}
\def\mov#1(#2,#3)#4{\rlap{\L*=#1\Lengthunit\kern#2\L*\raise#3\L*\hbox{#4}}}
\def\smov#1(#2,#3)#4{\rlap{\L*=#1\Lengthunit
\xL*=\xscale\L*\yL*=\yscale\L*\kern#2\xL*\raise#3\yL*\hbox{#4}}}
\def\mov*(#1,#2)#3{\rlap{\kern#1\raise#2\hbox{#3}}}
\def\lin#1(#2,#3){\rlap{\sm*\mov#1(#2,#3){\sl*}}}
\def\arr*(#1,#2,#3){\mov*(#1\dd*,#1\dt*){%
\sm*\mov*(#2\dd*,#2\dt*){\mov*(#3\dt*,-#3\dd*){\sl*}}%
\sm*\mov*(#2\dd*,#2\dt*){\mov*(-#3\dt*,#3\dd*){\sl*}}}}
\def\arrow#1(#2,#3){\rlap{\lin#1(#2,#3)\mov#1(#2,#3){%
\d**=-.012\Lengthunit\dd*=#2\d**\dt*=#3\d**%
\arr*(1,10,4)\arr*(3,8,4)\arr*(4.8,4.2,3)}}}
\def\arrlin#1(#2,#3){\rlap{\L*=#1\Lengthunit\L*=.5\L*%
\lin#1(#2,#3)\mov*(#2\L*,#3\L*){\arrow.1(#2,#3)}}}
\def\dasharrow#1(#2,#3){\rlap{%
{\Lengthunit=0.9\Lengthunit\dashlin#1(#2,#3)\mov#1(#2,#3){\sm*}}%
\mov#1(#2,#3){\sl*\d**=-.012\Lengthunit\dd*=#2\d**\dt*=#3\d**%
\arr*(1,10,4)\arr*(3,8,4)\arr*(4.8,4.2,3)}}}
\def\clap#1{\hbox to 0pt{\hss #1\hss}}
\def\ind(#1,#2)#3{\rlap{%
\d*=.1\Lengthunit\kern#1\d*\raise#2\d*\hbox{\lower2pt\clap{$#3$}}}}
\def\sh*(#1,#2)#3{\rlap{%
\dm*=\the\n*\d**\xL*=\xscale\dm*\yL*=\yscale\dm*
\kern#1\xL*\raise#2\yL*\hbox{#3}}}
\def\calcnum*#1(#2,#3){\a*=1000sp\b*=1000sp\a*=#2\a*\b*=#3\b*%
\ifdim\a*<0pt\a*-\a*\fi\ifdim\b*<0pt\b*-\b*\fi%
\ifdim\a*>\b*\c*=.96\a*\advance\c*.4\b*%
\else\c*=.96\b*\advance\c*.4\a*\fi%
\k*\a*\multiply\k*\k*\l*\b*\multiply\l*\l*%
\m*\k*\advance\m*\l*\n*\c*\r*\n*\multiply\n*\n*%
\dn*\m*\advance\dn*-\n*\divide\dn*2\divide\dn*\r*%
\advance\r*\dn*%
\c*=\the\Nhalfperiods5sp\c*=#1\c*\ifdim\c*<0pt\c*-\c*\fi%
\multiply\c*\r*\N*\c*\divide\N*10000}
\def\dashlin#1(#2,#3){\rlap{\calcnum*#1(#2,#3)%
\d**=#1\Lengthunit\ifdim\d**<0pt\d**-\d**\fi%
\divide\N*2\multiply\N*2\advance\N*1%
\divide\d**\N*\sm*\n*\*one\sh*(#2,#3){\sl*}%
\loop\advance\n*\*one\sh*(#2,#3){\sm*}\advance\n*\*one\sh*(#2,#3){\sl*}%
\ifnum\n*<\N*\repeat}}
\def\dashdotlin#1(#2,#3){\rlap{\calcnum*#1(#2,#3)%
\d**=#1\Lengthunit\ifdim\d**<0pt\d**-\d**\fi%
\divide\N*2\multiply\N*2\advance\N*1\multiply\N*2%
\divide\d**\N*\sm*\n*\*two\sh*(#2,#3){\sl*}\loop%
\advance\n*\*one\sh*(#2,#3){\kern-1.48pt\lower.5pt\hbox{\rm.}}%
\advance\n*\*one\sh*(#2,#3){\sm*}%
\advance\n*\*two\sh*(#2,#3){\sl*}\ifnum\n*<\N*\repeat}}
\def\shl*(#1,#2)#3{\kern#1#3\lower#2#3\hbox{\unhcopy\spl*}}
\def\trianglin#1(#2,#3){\rlap{\toks0={#2}\toks1={#3}\calcnum*#1(#2,#3)%
\dd*=.57\Lengthunit\dd*=#1\dd*\divide\dd*\N*%
\d**=#1\Lengthunit\ifdim\d**<0pt\d**-\d**\fi%
\multiply\N*2\divide\d**\N*\advance\N*-1\sm*\n*\*one\loop%
\shl**{\dd*}\dd*-\dd*\advance\n*2%
\ifnum\n*<\N*\repeat\n*\N*\advance\n*1\shl**{0pt}}}
\def\wavelin#1(#2,#3){\rlap{\toks0={#2}\toks1={#3}\calcnum*#1(#2,#3)%
\dd*=.23\Lengthunit\dd*=#1\dd*\divide\dd*\N*%
\d**=#1\Lengthunit\ifdim\d**<0pt\d**-\d**\fi%
\multiply\N*4\divide\d**\N*\sm*\n*\*one\loop%
\shl**{\dd*}\dt*=1.3\dd*\advance\n*1%
\shl**{\dt*}\advance\n*\*one%
\shl**{\dd*}\advance\n*\*two%
\dd*-\dd*\ifnum\n*<\N*\repeat\n*\N*\shl**{0pt}}}
\def\w*lin(#1,#2){\rlap{\toks0={#1}\toks1={#2}\d**=\Lengthunit\dd*=-.12\d**%
\N*8\divide\d**\N*\sm*\n*\*one\loop%
\shl**{\dd*}\dt*=1.3\dd*\advance\n*\*one%
\shl**{\dt*}\advance\n*\*one%
\shl**{\dd*}\advance\n*\*one%
\shl**{0pt}\dd*-\dd*\advance\n*1\ifnum\n*<\N*\repeat}}
\def\l*arc(#1,#2)[#3][#4]{\rlap{\toks0={#1}\toks1={#2}\d**=\Lengthunit%
\dd*=#3.037\d**\dd*=#4\dd*\dt*=#3.049\d**\dt*=#4\dt*\ifdim\d**>16mm%
\d**=.25\d**\n*\*one\shl**{-\dd*}\n*\*two\shl**{-\dt*}\n*3\relax%
\shl**{-\dd*}\n*4\relax\shl**{0pt}\else\ifdim\d**>5mm%
\d**=.5\d**\n*\*one\shl**{-\dt*}\n*\*two\shl**{0pt}%
\else\n*\*one\shl**{0pt}\fi\fi}}
\def\d*arc(#1,#2)[#3][#4]{\rlap{\toks0={#1}\toks1={#2}\d**=\Lengthunit%
\dd*=#3.037\d**\dd*=#4\dd*\d**=.25\d**\sm*\n*\*one\shl**{-\dd*}%
\n*3\relax\sh*(#1,#2){\xL*=\xscale\dd*\yL*=\yscale\dd*
\kern#2\xL*\lower#1\yL*\hbox{\sm*}}%
\n*4\relax\shl**{0pt}}}
\def\arc#1[#2][#3]{\rlap{\Lengthunit=#1\Lengthunit%
\sm*\l*arc(#2.1914,#3.0381)[#2][#3]%
\smov(#2.1914,#3.0381){\l*arc(#2.1622,#3.1084)[#2][#3]}%
\smov(#2.3536,#3.1465){\l*arc(#2.1084,#3.1622)[#2][#3]}%
\smov(#2.4619,#3.3086){\l*arc(#2.0381,#3.1914)[#2][#3]}}}
\def\dasharc#1[#2][#3]{\rlap{\Lengthunit=#1\Lengthunit%
\d*arc(#2.1914,#3.0381)[#2][#3]%
\smov(#2.1914,#3.0381){\d*arc(#2.1622,#3.1084)[#2][#3]}%
\smov(#2.3536,#3.1465){\d*arc(#2.1084,#3.1622)[#2][#3]}%
\smov(#2.4619,#3.3086){\d*arc(#2.0381,#3.1914)[#2][#3]}}}
\def\wavearc#1[#2][#3]{\rlap{\Lengthunit=#1\Lengthunit%
\w*lin(#2.1914,#3.0381)%
\smov(#2.1914,#3.0381){\w*lin(#2.1622,#3.1084)}%
\smov(#2.3536,#3.1465){\w*lin(#2.1084,#3.1622)}%
\smov(#2.4619,#3.3086){\w*lin(#2.0381,#3.1914)}}}
\def\shl**#1{\c*=\the\n*\d**\d*=#1%
\a*=\the\toks0\c*\b*=\the\toks1\d*\advance\a*-\b*%
\b*=\the\toks1\c*\d*=\the\toks0\d*\advance\b*\d*%
\a*=\xscale\a*\b*=\yscale\b*%
\raise\b*\rlap{\kern\a*\unhcopy\spl*}}
\def\wlin*#1(#2,#3)[#4]{\rlap{\toks0={#2}\toks1={#3}%
\c*=#1\l*\c*\c*=.01\Lengthunit\m*\c*\divide\l*\m*%
\c*=\the\Nhalfperiods5sp\multiply\c*\l*\N*\c*\divide\N*\*ths%
\divide\N*2\multiply\N*2\advance\N*1%
\dd*=.002\Lengthunit\dd*=#4\dd*\multiply\dd*\l*\divide\dd*\N*%
\d**=#1\multiply\N*4\divide\d**\N*\sm*\n*\*one\loop%
\shl**{\dd*}\dt*=1.3\dd*\advance\n*\*one%
\shl**{\dt*}\advance\n*\*one%
\shl**{\dd*}\advance\n*\*two%
\dd*-\dd*\ifnum\n*<\N*\repeat\n*\N*\shl**{0pt}}}
\def\wavebox#1{\setbox0\hbox{#1}%
\a*=\wd0\advance\a*14pt\b*=\ht0\advance\b*\dp0\advance\b*14pt%
\hbox{\kern9pt%
\mov*(0pt,\ht0){\mov*(-7pt,7pt){\wlin*\a*(1,0)[+]\wlin*\b*(0,-1)[-]}}%
\mov*(\wd0,-\dp0){\mov*(7pt,-7pt){\wlin*\a*(-1,0)[+]\wlin*\b*(0,1)[-]}}%
\box0\kern9pt}}
\def\rectangle#1(#2,#3){%
\lin#1(#2,0)\lin#1(0,#3)\mov#1(0,#3){\lin#1(#2,0)}\mov#1(#2,0){\lin#1(0,#3)}}
\def\dashrectangle#1(#2,#3){\dashlin#1(#2,0)\dashlin#1(0,#3)%
\mov#1(0,#3){\dashlin#1(#2,0)}\mov#1(#2,0){\dashlin#1(0,#3)}}
\def\waverectangle#1(#2,#3){\L*=#1\Lengthunit\a*=#2\L*\b*=#3\L*%
\ifdim\a*<0pt\a*-\a*\def\x*{-1}\else\def\x*{1}\fi%
\ifdim\b*<0pt\b*-\b*\def\y*{-1}\else\def\y*{1}\fi%
\wlin*\a*(\x*,0)[-]\wlin*\b*(0,\y*)[+]%
\mov#1(0,#3){\wlin*\a*(\x*,0)[+]}\mov#1(#2,0){\wlin*\b*(0,\y*)[-]}}
\def\calcparab*{%
\ifnum\n*>\m*\k*\N*\advance\k*-\n*\else\k*\n*\fi%
\a*=\the\k* sp\a*=10\a*\b*\dm*\advance\b*-\a*\k*\b*%
\a*=\the\*ths\b*\divide\a*\l*\multiply\a*\k*%
\divide\a*\l*\k*\*ths\r*\a*\advance\k*-\r*%
\dt*=\the\k*\L*}
\def\arcto#1(#2,#3)[#4]{\rlap{\toks0={#2}\toks1={#3}\calcnum*#1(#2,#3)%
\dm*=135sp\dm*=#1\dm*\d**=#1\Lengthunit\ifdim\dm*<0pt\dm*-\dm*\fi%
\multiply\dm*\r*\a*=.3\dm*\a*=#4\a*\ifdim\a*<0pt\a*-\a*\fi%
\advance\dm*\a*\N*\dm*\divide\N*10000%
\divide\N*2\multiply\N*2\advance\N*1%
\L*=-.25\d**\L*=#4\L*\divide\d**\N*\divide\L*\*ths%
\m*\N*\divide\m*2\dm*=\the\m*5sp\l*\dm*%
\sm*\n*\*one\loop\calcparab*\shl**{-\dt*}%
\advance\n*1\ifnum\n*<\N*\repeat}}
\def\arrarcto#1(#2,#3)[#4]{\L*=#1\Lengthunit\L*=.54\L*%
\arcto#1(#2,#3)[#4]\mov*(#2\L*,#3\L*){\d*=.457\L*\d*=#4\d*\d**-\d*%
\mov*(#3\d**,#2\d*){\arrow.02(#2,#3)}}}
\def\dasharcto#1(#2,#3)[#4]{\rlap{\toks0={#2}\toks1={#3}\calcnum*#1(#2,#3)%
\dm*=\the\N*5sp\a*=.3\dm*\a*=#4\a*\ifdim\a*<0pt\a*-\a*\fi%
\advance\dm*\a*\N*\dm*%
\divide\N*20\multiply\N*2\advance\N*1\d**=#1\Lengthunit%
\L*=-.25\d**\L*=#4\L*\divide\d**\N*\divide\L*\*ths%
\m*\N*\divide\m*2\dm*=\the\m*5sp\l*\dm*%
\sm*\n*\*one\loop%
\calcparab*\shl**{-\dt*}\advance\n*1%
\ifnum\n*>\N*\else\calcparab*%
\sh*(#2,#3){\kern#3\dt*\lower#2\dt*\hbox{\sm*}}\fi%
\advance\n*1\ifnum\n*<\N*\repeat}}
\def\*shl*#1{%
\c*=\the\n*\d**\advance\c*#1\a**\d*\dt*\advance\d*#1\b**%
\a*=\the\toks0\c*\b*=\the\toks1\d*\advance\a*-\b*%
\b*=\the\toks1\c*\d*=\the\toks0\d*\advance\b*\d*%
\raise\b*\rlap{\kern\a*\unhcopy\spl*}}
\def\calcnormal*#1{%
\b**=10000sp\a**\b**\k*\n*\advance\k*-\m*%
\multiply\a**\k*\divide\a**\m*\a**=#1\a**\ifdim\a**<0pt\a**-\a**\fi%
\ifdim\a**>\b**\d*=.96\a**\advance\d*.4\b**%
\else\d*=.96\b**\advance\d*.4\a**\fi%
\d*=.01\d*\r*\d*\divide\a**\r*\divide\b**\r*%
\ifnum\k*<0\a**-\a**\fi\d*=#1\d*\ifdim\d*<0pt\b**-\b**\fi%
\k*\a**\a**=\the\k*\dd*\k*\b**\b**=\the\k*\dd*}
\def\wavearcto#1(#2,#3)[#4]{\rlap{\toks0={#2}\toks1={#3}\calcnum*#1(#2,#3)%
\c*=\the\N*5sp\a*=.4\c*\a*=#4\a*\ifdim\a*<0pt\a*-\a*\fi%
\advance\c*\a*\N*\c*\divide\N*20\multiply\N*2\advance\N*-1\multiply\N*4%
\d**=#1\Lengthunit\dd*=.012\d**\ifdim\d**<0pt\d**-\d**\fi\L*=.25\d**%
\divide\d**\N*\divide\dd*\N*\L*=#4\L*\divide\L*\*ths%
\m*\N*\divide\m*2\dm*=\the\m*0sp\l*\dm*%
\sm*\n*\*one\loop\calcnormal*{#4}\calcparab*%
\*shl*{1}\advance\n*\*one\calcparab*%
\*shl*{1.3}\advance\n*\*one\calcparab*%
\*shl*{1}\advance\n*2%
\dd*-\dd*\ifnum\n*<\N*\repeat\n*\N*\shl**{0pt}}}
\def\triangarcto#1(#2,#3)[#4]{\rlap{\toks0={#2}\toks1={#3}\calcnum*#1(#2,#3)%
\c*=\the\N*5sp\a*=.4\c*\a*=#4\a*\ifdim\a*<0pt\a*-\a*\fi%
\advance\c*\a*\N*\c*\divide\N*20\multiply\N*2\advance\N*-1\multiply\N*2%
\d**=#1\Lengthunit\dd*=.012\d**\ifdim\d**<0pt\d**-\d**\fi\L*=.25\d**%
\divide\d**\N*\divide\dd*\N*\L*=#4\L*\divide\L*\*ths%
\m*\N*\divide\m*2\dm*=\the\m*0sp\l*\dm*%
\sm*\n*\*one\loop\calcnormal*{#4}\calcparab*%
\*shl*{1}\advance\n*2%
\dd*-\dd*\ifnum\n*<\N*\repeat\n*\N*\shl**{0pt}}}
\def\hr*#1{\clap{\xL*=\xscale\Lengthunit\vrule width#1\xL* height.1pt}}
\def\shade#1[#2]{\rlap{\Lengthunit=#1\Lengthunit%
\smov(0,#2.05){\hr*{.994}}\smov(0,#2.1){\hr*{.980}}%
\smov(0,#2.15){\hr*{.953}}\smov(0,#2.2){\hr*{.916}}%
\smov(0,#2.25){\hr*{.867}}\smov(0,#2.3){\hr*{.798}}%
\smov(0,#2.35){\hr*{.715}}\smov(0,#2.4){\hr*{.603}}%
\smov(0,#2.45){\hr*{.435}}}}
\def\dshade#1[#2]{\rlap{%
\Lengthunit=#1\Lengthunit\if#2-\def\t*{+}\else\def\t*{-}\fi%
\smov(0,\t*.025){%
\smov(0,#2.05){\hr*{.995}}\smov(0,#2.1){\hr*{.988}}%
\smov(0,#2.15){\hr*{.969}}\smov(0,#2.2){\hr*{.937}}%
\smov(0,#2.25){\hr*{.893}}\smov(0,#2.3){\hr*{.836}}%
\smov(0,#2.35){\hr*{.760}}\smov(0,#2.4){\hr*{.662}}%
\smov(0,#2.45){\hr*{.531}}\smov(0,#2.5){\hr*{.320}}}}}
\def\vdot{\rlap{\kern-1.9pt\lower1.8pt\hbox{$\scriptstyle\bullet$}}}
\def\vtimes{\rlap{\kern-3pt\lower1.8pt\hbox{$\scriptstyle\times$}}}
\def\vDot{\rlap{\kern-2.3pt\lower2.7pt\hbox{$\bullet$}}}
\def\vTimes{\rlap{\kern-3.6pt\lower2.4pt\hbox{$\times$}}}
\catcode`\*=12
\newcount\CatcodeOfAtSign
\CatcodeOfAtSign=\the\catcode`\@
\catcode`\@=11
\newcount\n@ast
\def\n@ast@#1{\n@ast0\relax\get@ast@#1\end}
\def\get@ast@#1{\ifx#1\end\let\next\relax\else%
\ifx#1*\advance\n@ast1\fi\let\next\get@ast@\fi\next}
\newif\if@up \newif\if@dwn
\def\up@down@#1{\@upfalse\@dwnfalse%
\if#1u\@uptrue\fi\if#1U\@uptrue\fi\if#1+\@uptrue\fi%
\if#1d\@dwntrue\fi\if#1D\@dwntrue\fi\if#1-\@dwntrue\fi}
\def\halfcirc#1(#2)[#3]{{\Lengthunit=#2\Lengthunit\up@down@{#3}%
\if@up\smov(0,.5){\arc[-][-]\arc[+][-]}\fi%
\if@dwn\smov(0,-.5){\arc[-][+]\arc[+][+]}\fi%
\def\lft{\smov(0,.5){\arc[-][-]}\smov(0,-.5){\arc[-][+]}}%
\def\rght{\smov(0,.5){\arc[+][-]}\smov(0,-.5){\arc[+][+]}}%
\if#3l\lft\fi\if#3L\lft\fi\if#3r\rght\fi\if#3R\rght\fi%
\n@ast@{#1}%
\ifnum\n@ast>0\if@up\shade[+]\fi\if@dwn\shade[-]\fi\fi%
\ifnum\n@ast>1\if@up\dshade[+]\fi\if@dwn\dshade[-]\fi\fi}}
\def\halfdashcirc(#1)[#2]{{\Lengthunit=#1\Lengthunit\up@down@{#2}%
\if@up\smov(0,.5){\dasharc[-][-]\dasharc[+][-]}\fi%
\if@dwn\smov(0,-.5){\dasharc[-][+]\dasharc[+][+]}\fi%
\def\lft{\smov(0,.5){\dasharc[-][-]}\smov(0,-.5){\dasharc[-][+]}}%
\def\rght{\smov(0,.5){\dasharc[+][-]}\smov(0,-.5){\dasharc[+][+]}}%
\if#2l\lft\fi\if#2L\lft\fi\if#2r\rght\fi\if#2R\rght\fi}}
\def\halfwavecirc(#1)[#2]{{\Lengthunit=#1\Lengthunit\up@down@{#2}%
\if@up\smov(0,.5){\wavearc[-][-]\wavearc[+][-]}\fi%
\if@dwn\smov(0,-.5){\wavearc[-][+]\wavearc[+][+]}\fi%
\def\lft{\smov(0,.5){\wavearc[-][-]}\smov(0,-.5){\wavearc[-][+]}}%
\def\rght{\smov(0,.5){\wavearc[+][-]}\smov(0,-.5){\wavearc[+][+]}}%
\if#2l\lft\fi\if#2L\lft\fi\if#2r\rght\fi\if#2R\rght\fi}}
\def\Circle#1(#2){\halfcirc#1(#2)[u]\halfcirc#1(#2)[d]\n@ast@{#1}%
\ifnum\n@ast>0\clap{%
\dimen0=\xscale\Lengthunit\vrule width#2\dimen0 height.1pt}\fi}
\def\wavecirc(#1){\halfwavecirc(#1)[u]\halfwavecirc(#1)[d]}
\def\dashcirc(#1){\halfdashcirc(#1)[u]\halfdashcirc(#1)[d]}
%
\def\xscale{1}
\def\yscale{1}
\def\Ellipse#1(#2)[#3,#4]{\def\xscale{#3}\def\yscale{#4}%
\Circle#1(#2)\def\xscale{1}\def\yscale{1}}
\def\dashEllipse(#1)[#2,#3]{\def\xscale{#2}\def\yscale{#3}%
\dashcirc(#1)\def\xscale{1}\def\yscale{1}}
\def\waveEllipse(#1)[#2,#3]{\def\xscale{#2}\def\yscale{#3}%
\wavecirc(#1)\def\xscale{1}\def\yscale{1}}
\def\halfEllipse#1(#2)[#3][#4,#5]{\def\xscale{#4}\def\yscale{#5}%
\halfcirc#1(#2)[#3]\def\xscale{1}\def\yscale{1}}
\def\halfdashEllipse(#1)[#2][#3,#4]{\def\xscale{#3}\def\yscale{#4}%
\halfdashcirc(#1)[#2]\def\xscale{1}\def\yscale{1}}
\def\halfwaveEllipse(#1)[#2][#3,#4]{\def\xscale{#3}\def\yscale{#4}%
\halfwavecirc(#1)[#2]\def\xscale{1}\def\yscale{1}}
\catcode`\@=\the\CatcodeOfAtSign

\title{Three-dimensional Lorentz-violating action}

\author{J. R. Nascimento}
\email{jroberto@fisica.ufpb.br}
\affiliation{Departamento de F\'\i sica, Universidade Federal da Para\'\i ba, 
Caixa Postal 5008, 58051-970 Jo\~ao Pessoa, Para\'\i ba, Brazil}
\author{A. Yu. Petrov}
\email{petrov@fisica.ufpb.br}
\affiliation{Departamento de F\'\i sica, Universidade Federal da Para\'\i ba, 
Caixa Postal 5008, 58051-970 Jo\~ao Pessoa, Para\'\i ba, Brazil}
\author{C. Wotzasek}
\email{clovis@if.ufrj.br}
\affiliation{Instituto de F\'\i sica, Universidade Federal de Rio de Janeiro, 
Caixa Postal 21945, Rio de Janeiro, Brazil}
\author{C. A. D. Zarro}
\email{carlos.zarro@if.ufrj.br}
\affiliation{Instituto de F\'\i sica, Universidade Federal de Rio de Janeiro, 
Caixa Postal 21945, Rio de Janeiro, Brazil}

\begin{abstract}
We demonstrate the generation of the three-dimensional Chern-Simons-like Lorentz-breaking ``mixed" quadratic action via an appropriate Lorentz-breaking coupling of vector and scalar fields to the spinor field and study some features of the scalar QED with such a term. We show that the same term emerges through a nonpertubative method, namely the Julia-Toulouse approach of condensation of charges  and defects.
\end{abstract}

\maketitle
\newpage

\section{Introduction}

The Lorentz symmetry breaking is intensively studied now (for some observational results see \cite{Bert}). One of the most interesting lines of its investigation consists in constructing the Lorentz-breaking extensions of the known physical models. First description of the possibilities for these extensions was carried out in \cite{Kost}. Further, many examples of the new Lorentz-breaking terms were generated due to appropriate couplings of scalar, spinor and gravitational fields with the spinor ones. The most important examples of such terms are, first, the four-dimensional Lorentz-breaking Chern-Simons-like term originally introduced by Jackiw and collaborators \cite{JK}, second, the non-Abelian generalization of this term \cite{YM4d,ptime}, third, the gravitational Chern-Simons term \cite{JaPi,ptime}. We can note also other manners of description of the Lorentz symmetry breaking such as noncommutativity \cite{NC} and double special relativity \cite{DSR}.

However, all these results are four-dimensional ones. At the same time, the three-dimensional space-time represents itself as a convenient laboratory for study of many physical effects. The main reasons for it are the more simpler form and one-loop finiteness for almost all field theory models. The main results achieved in study of the Lorentz symmetry breaking in three-dimensional space-time are, first, generation of many Lorentz-breaking terms as a consequence of the spontaneous Lorentz symmetry breaking in the three-dimensional bumblebee model through a tadpole method with use of the reducible representation of the Dirac matrices \cite{Char}, second, generalization of duality \cite{dual0} for the Lorentz-breaking models implying in arising of new couplings between scalar, spinor and gauge fields \cite{dual}, third, obtaining of new terms via dimensional reduction of the electrodynamics with the four-dimensional Lorentz-breaking Chern-Simons-like term \cite{Ferr}. In all these papers, a new 
mixed quadratic term involving both scalar and electromagnetic fields was shown to arise. Some possible applications of this term within the confinement context were discussed in \cite{JT,JTArtigao}.  Therefore, the very natural question consists in possibility of generating this term through more simple and traditional mechanisms of the Lorentz-breaking couplings of scalar and gauge fields to the spinor one which could be similar to \cite{YM4d,ptime}, and through the Julia-Toulouse approach \cite{JT,JTArtigao}. 

The Julia-Toulouse approach (JTA) consists in a prescription to obtain a low-energy effective field theory describing a system where a condensation of topological currents has occurred. Initially, these topological currents are sparsely distributed through the system constituting the diluted phase. Then, there is a proliferation of topological currents due to a condensation mechanism that is beyond the scope of JTA since in the Julia-Toulouse method the condensation process is taken for granted. The original proposal of this technique was done in the realm of condensed matter physics in Ref. \cite{JTOriginal}, latter, this procedure was generalized to relativistic quantum fields in Ref. \cite{JTQuevedo}. The original JTA relies on duality transformations, since to apply the JTA the first step is to get the dual theory on the diluted phase before applying the prescription and then obtain the effective theory on the condensed regime on a dual theory. The final step consists in dualize again to finally 
find the the effective field theory of the original condensed phase. However, this original procedure which depends on duality tranformations can sometimes be cumbersome and indeed it is not necessary as  shown in Refs. \cite{JT,JTArtigao}. This new procedure, dubbed Generalized Julia Toulouse Approach (GJTA), is based on the JT rationale  and its cornerstone uses the generalized Poisson identity. This identity makes clear the physical content of the condensation of topological currents and this avoids the two dual transformations to implement the original JTA. Another advantage of GJTA is that it can be applied to models that do not admit a dual theory \cite{Guimaraes:2012tx}. For a comprehensive discussion of GJTA with applications, the reader is referred to \cite{JTArtigao}.

This manuscript is organized as follows: In Sec. \ref{Sec:Perturbative}, two different forms to get the ``mixed'' term are presented: one via Feynman diagrams methods (Sec. \ref{Sec:PerturbativeFeynman}) and using proper-time approach (Sec. \ref{Sec:PerturbativeSchwinger}). In Sec. \ref{Sec:GJTA}, the GJTA is briefly presented and it is used to obtain the same ``mixed'' term as before. The conclusions are present in Sec. \ref{Sec:Conclusions} and the corrections  on the physical spectra due to the ``mixed'' term are given in the Appendix \ref{Sec:Application}.

\section{Perturbative approach}\label{Sec:Perturbative}

\subsection{Feynman Diagram Methods} \label{Sec:PerturbativeFeynman}

Let us consider the model of fermions interacting with scalar field $\phi(x)$ and vector one $A_{\mu}(x)$, with the Lorentz symmetry violation is implemented via a constant vector $a^{\mu}$. We consider the Lagrangian involving the Lorentz-breaking generalization of the Yukawa coupling \cite {aether} (we note that this coupling is renormalizable, see discussion of the renormalizability of the Lorentz-breaking theories in \cite{ABPS}):
\bea
\label{l}
{\cal L}_{f}=-\frac{1}{4}F_{\mu\nu}F^{\mu\nu}-\frac{1}{2}\phi(\Box+M^2)\phi+\bar{\psi}\left(i \ds- m -  e\As-  g\as \phi  \right)\psi.
\eea
We note that, unlike the four-dimensional theory (see f.e. \cite{YM4d}) where the Lorentz symmetry breaking has been introduced through an additive term $\bs\gamma_5$, with $b^{\mu}$ is a Lorentz-breaking pseudo-vector, 
in three dimensions this manner of implementing the Lorentz symmetry breaking is the most adequate one since the $\gamma_5$ matrix now is simply an unit matrix, thus, the impact of this additive term can be completely removed through an appropriate redefinition of the $A^{\mu}$ field.
Integrating out the spinor fields, we arrive at their following complete one-loop effective action:
\bea
\label{trace}
\Gamma^{(1)}=i{\rm Tr}\ln\left(i \ds- m -  e\As-  g\as \phi  \right).
\eea
Within this paper, our aim consists in calculating the one-loop Chern-Simons-like mixed effective action of the form 
\bea
\label{jt}
\Gamma=\int d^3x\epsilon^{\mu\nu\lambda}a_{\mu}F_{\nu\lambda}\phi.
\eea
Some issues related to this effective action were discussed in \cite{Char,dual,Ferr,JT, JTArtigao}.
It is natural to suggest that in the momentum space it can be represented as
\bea
\Gamma=\int\frac{d^3q}{(2\pi)^3}\phi(-q)\Pi^{\mu}(q)A_{\mu}(q),
\eea
with the $\Pi^{\mu}(q)$ is the self-energy tensor.
We note that in this theory also other quadratic contributions to the action are generated, for example, the Chern-Simons term; however, here we concentrate only on the mixed term (\ref{jt}).

Applying the following Feynman rules

\vspace*{2mm}

\hspace{2cm}
\Lengthunit=1.2cm
\GRAPH(hsize=3){\dashlin(1,0)\ind(15,0){\;\;\;\;\;\;\;\;\;\;\;\;=\frac{i(\ps+m)}{p^{2}-m^{2}}}
}
\hspace{.5cm}
\Lengthunit=1.2cm
\GRAPH(hsize=3){\dashlin(1,0)\mov(.5,0){\wavelin(0,.7)}\ind(22,0){=-ie\gamma^{\mu}}
}
\hspace{.5cm}
\Lengthunit=1.2cm
\GRAPH(hsize=3){\ind(5,0){\bullet}\dashlin(1,0)\mov(.5,0){\lin(0,.7)}\ind(22,0){=-ig\as}
}

\vspace*{2mm}

\noindent where the dot denotes the Lorentz-breaking insertion in the vertex, we arrive at the following diagram which contributes to the two-point ``mixed" function of the scalar and vector fields: 

\vspace*{3mm}

\hspace{4.0cm}
\Lengthunit=1.2cm
\GRAPH(hsize=2){\wavelin(.5,0)\mov(1,0){\dashcirc(1)}\mov(1.5,0){\lin(.5,0)}\ind(15,0){\bullet}
}
\vspace*{3mm}

\noindent Here the dashed line is for the propagator of the $\psi$ field, the wavy line -- for the external $A_{\mu}$ field, and the single line -- for the external $\phi$ field.

The contribution of this diagram evidently looks like
\bea
I=-eg{\rm Tr}\int\frac{d^3p}{(2\pi)^3}\int \frac{d^3k}{(2\pi)^3}\As(-p)(\ks+m)\phi(p)\as(\ks+\ps+m)\frac{1}{(k^2-m^2)[(k+p)^2-m^2]}.
\eea
To obtain the term (\ref{jt}) proportional to the Levi-Civita symbol we must take into account the products of three Dirac matrices only:
\bea
I=-egm{\rm Tr}\int\frac{d^3p}{(2\pi)^3}A^{\mu}(-p)\phi(p)a^{\nu}\int \frac{d^3k}{(2\pi)^3}\frac{\gamma_{\mu}\gamma_{\alpha}\gamma_{\nu}k^{\alpha}+\gamma_{\mu}\gamma_{\nu}\gamma_{\alpha}(k^{\alpha}+p^{\alpha})}{(k^2-m^2)[(k+p)^2-m^2]}.
\eea
We choose the signature $diag(+--)$, the corresponding Dirac matrices are: $(\gamma^0)^{\alpha}_{\phantom{\alpha}\beta}=\sigma^2, (\gamma^1)^{\alpha}_{\phantom{\alpha}\beta}=i\sigma^1, (\gamma^0)^{\alpha}_{\phantom{\alpha}\beta}=i\sigma^3$, they satisfy relations: $\{\gamma^{\mu},\gamma^{\nu}\}=2\eta^{\mu\nu}$, ${\rm tr}(\gamma^{\mu}\gamma^{\nu}\gamma^{\lambda})=2i\epsilon^{\mu\nu\lambda}$. Using these relations, we can simplify the expression for the contribution above:
\bea
I=-2i\epsilon_{\alpha\mu\nu}egm\int\frac{d^3p}{(2\pi)^3}p^{\alpha}A^{\mu}(-p)\phi(p)a^{\nu}\int \frac{d^3k}{(2\pi)^3}\frac{1}{(k^2-m^2)[(k+p)^2-m^2]}.
\eea
After Wick rotation and integration over momenta we arrive at
\bea
I=\epsilon_{\alpha\mu\nu}eg\frac{m}{4\pi|m|}\int\frac{d^3p}{(2\pi)^3}p^{\alpha}A^{\mu}(-p)\phi(p)a^{\nu}.
\eea
Carrying out the inverse Wick rotation and inverse Fourier transform, we find after some simple transformations
\bea
I=-eg\frac{m}{8\pi|m|}\int d^3x \epsilon^{\alpha\mu\nu}F_{\alpha\mu}a_{\nu}\phi.
\eea
This is a desired ``mixed" term (\ref{jt}). It possesses restricted gauge invariance (cf. \cite{dual}), i.e. it is invariant under the gauge transformations $\delta A_{\mu}=\partial_{\mu}\xi$, with the scalar $\phi$ stays untouched. We note that the dependence of this result on the sign of the mass $m$ originates from the ambiguity of choice of the direction of the Lorentz-breaking vector $a^{\mu}$ \cite{Grig}.

\subsection{The Schwinger Proper-time Method}\label{Sec:PerturbativeSchwinger}
Alternatively, we can also calculate the same term via the proper-time method. To do it, we study the expression (\ref{trace}). First, we can rewrite this expression in the form ${\rm Tr}\ln(\Box+{\cal M})$, adding to the right-hand side of (\ref{trace}) a constant $i{\rm Tr}\ln\left(i \ds+ m\right)$, similarly to the \cite{ptime}.
As a result, the one-loop effective action (\ref{trace}) takes the form
\bea
\Gamma^{(1)}=i{\rm Tr}\ln(-\Box-m^2-e\As(i\ds+m)-g\phi \as(i\ds+m)), 
\eea
We can expand this expression up to the first order in $a^{\mu}$ which looks like
\bea
\Gamma^{(1)}_1=ig{\rm Tr}\left[\Big[\Box+m^2+e\As(i\ds+m)\Big]^{-1}\phi \as(i\ds+m)\right].
\eea
Now, we can use the Schwinger proper-time representation  $A^{-1}=i\int_0^{\infty}e^{isA}ds $:
\bea
\Gamma^{(1)}_1=-g{\rm Tr}\Big[\int_0^{\infty} ds e^{is(\Box+m^2+e\As(i\ds+m))}\phi \as(i\ds+m)\Big].
\eea
To evaluate the exponential, we use the Hausdorf formula whose form sufficient in our case is $e^{A+B}=e^Ae^Be^{-\frac{[A,B]}{2}}$.
Thus, keeping into account only first derivatives of the $A_{\mu}$ and using cyclic property of the trace, we find
\bea
\Gamma^{(1)}_1=-g{\rm Tr}\Big[\int_0^{\infty} ds e^{ism^2}e^{ise\As(i\ds+m)}e^{-es^2(\pa^{\mu}\As)(i\ds+m)\pa_{\mu}}\phi \as(i\ds+m)e^{is\Box}\Big].
\eea
The derivatives act on all on the right. Now, we can keep in this expression only the first order in $A_{\mu}$:
\bea
\Gamma^{(1)}_1=-eg{\rm Tr}\Big[\int_0^{\infty} ds e^{ism^2}\Big(is\As(i\ds+m)-s^2(\pa^{\mu}\As)(i\ds+m)\pa_{\mu}\Big)\phi \as(i\ds+m)e^{is\Box}\Big].
\eea
It remains to calculate a trace. To obtain a desired term, we must take into account only contributions involving exactly three Dirac matrices and involving an even number of the derivatives acting on $e^{is\Box}$. As ${\rm tr}(\gamma^{\mu}\gamma^{\nu}\gamma^{\lambda})=2i\epsilon^{\mu\nu\lambda}$, we arrive at
\bea
\Gamma^{(1)}_1=-2egm\int d^3x\int_0^{\infty} ds s e^{ism^2}\epsilon^{\mu\nu\lambda}A_{\mu}(\pa_{\nu}\phi) a_{\lambda}e^{is\Box}\delta^3(x-x')|_{x=x'}.
\eea
After Fourier transform and Wick rotation we arrive at
\bea
\Gamma^{(1)}_1=-2egm\int d^3x\int \frac{d^3k}{(2\pi)^3}\int ds s e^{-sm^2}\epsilon^{\mu\nu\lambda}A_{\mu}(\pa_{\nu}\phi) a_{\lambda}e^{-sk^2}.
\eea
Calculation of the integrals over momenta and, then, over $s$ is straightforward, and we again arrive at
\bea
\label{1loop}
I=-eg\frac{m}{8\pi|m|}\int d^3x \epsilon^{\alpha\mu\nu}F_{\alpha\mu}a_{\nu}\phi.
\eea
This result is identically the same one obtained using the Feynman diagram approach. It is very natural since this contribution is superficially finite and hence does not involve any ambiguities.

\section{Lorentz-breaking mixed term and the Julia-Toulouse approach} \label{Sec:GJTA}

 In the previous sections we have showed that the ``mixed'' quadratic term can be successfully generated within the traditional perturbative approach. In this section, we show how the same term can be generated within an alternative, nonperturbative technique, that is, the Julia-Toulouse method.

To proceed with the Julia-Toulouse approach \cite{JTArtigao}, we start with the Lagrangian (\ref{l}), at the zero mass, and introduce the corresponding generating functional in the diluted phase:
\bea 
Z_d[j^{\mu}]=\int D A^{\mu} D\phi \exp \left[-i\int d^3x\left(
-\frac{1}{4}F_{\mu\nu}F^{\mu\nu}-\frac{1}{2}\phi\Box\phi+(-eA_{\mu}+g\phi a_{\mu})j^{\mu}
\right)\right],
\eea
where we absorbed the fermionic coupling into the current $j^{\mu}$. We choose the current to be of the form $j^{\mu}=\epsilon^{\mu\nu\alpha}\pa_{\nu}\chi_{\alpha}$, to be topologically conserved. The vector $\chi_{\alpha}$ is called the Chern kernel \cite{JTArtigao}. Then, following \cite{JTArtigao}, we  add a so-called activation term $\int d^3x\frac{j^{\mu}j_{\mu}}{2\Lambda}$ to the classical action (that is, the argument of the exponential) to introduce a defect condensation.  The parameter $\Lambda$ is related to the density of the condensate. It is a free parameter of the procedure and it can be fixed after comparing the effective field theory obtained by the JTA to the same theory computed by other methods \cite{JT,JTArtigao,JTQuevedo,Grigorio:2012jt,Guimaraes:2012ma}.  In particular, for the three-dimensional QED with magnetic monopoles, this parameter is fixed to maintain the consistency of this theory \cite{JT,JTArtigao}. Thus, the generating functional is modified, and we arrive at 
the new generation functional $Z_c$ describing the condensed phase:
\bea 
Z_c[j^{\mu}]&=&\sum\limits_{\{\chi_{\alpha}\}}\int D A^{\mu} D\phi \exp \left[-i\int d^3x\left(
-\frac{1}{4}F_{\mu\nu}F^{\mu\nu}-\frac{1}{2}\phi\Box\phi+(-eA_{\mu}+g\phi a_{\mu})\epsilon^{\mu\nu\alpha}\pa_{\nu}\chi_{\alpha}+ \right.\right.\nonumber\\
&+&\left.\left.
\frac{\epsilon^{\mu\nu\alpha}\pa_{\nu}\chi_{\alpha}\epsilon_{\mu\lambda\beta}\pa^{\lambda}\chi^{\beta}}{2\Lambda} \right)
\right],
\eea
Here we suggest the formal sum over the branes $\chi_{\alpha}$.
Let us promote their condensation. During this process, they convert to a vector field $B_{\alpha}$, which is formally described by introducing the integral over $B_{\alpha}$ and the functional delta function $\delta(\chi_{\alpha}-B_{\alpha})$, so, we have
\bea 
Z_c[j^{\mu}]&=&\sum\limits_{\{\chi_{\alpha}\}}\int D A^{\mu} D\phi DB^{\alpha}\delta(\chi_{\alpha}-B_{\alpha})
\exp \Big[-i\int d^3x\Big(
-\frac{1}{4}F_{\mu\nu}F^{\mu\nu}-\frac{1}{2}\phi\Box\phi+\nonumber\\&+&(-eA_{\mu}+g\phi a_{\mu})\epsilon^{\mu\nu\alpha}\pa_{\nu}\chi_{\alpha}+
\frac{\epsilon^{\mu\nu\alpha}\pa_{\nu}\chi_{\alpha}\epsilon_{\mu\lambda\beta}\pa^{\lambda}\chi^{\beta}}{2\Lambda}
\Big)\Big],
\eea
which is equivalent to
\bea 
Z_c[j^{\mu}]&=&\sum\limits_{\{\chi_{\alpha}\}}\int D A^{\mu} D\phi DB^{\alpha}\delta(\chi_{\alpha}-B_{\alpha})\exp \Big[-i\int d^3x\Big(
-\frac{1}{4}F_{\mu\nu}F^{\mu\nu}-\frac{1}{2}\phi\Box\phi+\nonumber\\&+&(-eA_{\mu}+g\phi a_{\mu})\epsilon^{\mu\nu\alpha}\pa_{\nu}B_{\alpha}+ 
\frac{\epsilon^{\mu\nu\alpha}\pa_{\nu}B_{\alpha}\epsilon_{\mu\lambda\beta}\pa^{\lambda}B^{\beta}}{2\Lambda}
\Big)\Big],
\eea
where the sum is taken over the branes.
Then we use a generalized Poisson identity \cite{JTArtigao}:
\bea
\sum\limits_{\{\chi_{\alpha}\}}\int DB^{\alpha}\delta(\chi_{\alpha}-B_{\alpha})=\sum\limits_{\{\Omega_{\mu\nu}\}}\exp(2\pi i\int d^3x\epsilon^{\mu\nu\rho}\Omega_{\mu\nu}B_{\rho}),
\eea
where $\Omega_{\mu\nu}$ is a magnetic vortex over the condensate \cite{JTArtigao}
and arrive at
\bea 
\label{zc}
Z_c[j^{\mu}]&=&\sum\limits_{\{\Omega_{\mu\nu}\}}\int D A^m D\phi DB^{\alpha}\exp \Big[-i\int d^3x\Big(
-\frac{1}{4}F_{\mu\nu}F^{\mu\nu}-\frac{1}{2}\phi\Box\phi+\nonumber\\&+&(-eA_{\mu}+g\phi a_{\mu})\epsilon^{\mu\nu\alpha}\pa_{\nu}B_{\alpha}+ 
\frac{\epsilon^{\mu\nu\alpha}\pa_{\nu}B_{\alpha}\epsilon_{\mu\lambda\beta}\pa^{\lambda}B^{\beta}}{2\Lambda}+ 2\pi \epsilon^{\mu\nu\rho}\Omega_{\mu\nu}B_{\rho}
\Big)\Big],
\eea
It remains only to integrate over the field $A_{\mu}$. Since it is gauge invariant, we add to the argument of the exponential the Feynman gauge fixing term $-\frac{1}{2}(\partial_{\mu}A^{\mu})^2$, after which the integral over $A_{\mu}$ turns out to be straightforward, by the rule
\bea
\label{rule}
\int DA^{\mu}\exp(i(-\frac{1}{2}A_{\mu}\Box A^{\mu}+A_{\nu}j^{\nu}))=\exp(i(\frac{1}{2}j_{\mu}\Box^{-1}j^{\mu}))
\eea
Then, we redefine $B_{\mu}\to \sqrt{\Lambda}B_{\mu}$ and arrive at
\bea 
Z_c[j^{\mu}]&=&\sum\limits_{\{\Omega_{\mu\nu}\}}\int D\phi DB^{\alpha}\exp \Big[-i\int d^3x\Big(
-\frac{1}{4}F_{\mu\nu}[B](\frac{e^2\Lambda}{\Box}-1)F^{\mu\nu}[B]-\frac{1}{2}\phi\Box\phi+\nonumber\\&+&g\phi\sqrt{\Lambda}  a_{\mu}\epsilon^{\mu\nu\alpha}F_{\nu\alpha}[B] 
+ 2\pi \epsilon^{\mu\nu\rho}\Omega_{\mu\nu}B_{\rho}
\Big)\Big],
\eea
where $F_{\mu\nu}[B]=\pa_{\mu}B_{\nu}-\pa_{\nu}B_{\mu}$. The last term vanishes since one considers the phase where the magnetic vortices are absent which represents a complete condensed phase. Notice, also that the term $-\frac{1}{4}F_{\mu\nu}[B]\frac{e^2\Lambda}{\Box}F^{\mu\nu}[B]$  represents a gauge invariant mass term for $B^{\mu}$ which can be seen straightforwardly performing integration by parts. This mass generating mechanism is a signature of the JTA. Hence, we suceeed generated the ``mixed'' term $g\phi\sqrt{\Lambda}a_{\mu}\epsilon^{\mu\nu\alpha}F_{\nu\alpha}[B]$ via GJTA. Interestingly, for the four-dimensional QED with Lorentz breaking, a very similar term, the Carrol-Field-Jackiw term, is induced by GJTA \cite{JTArtigao}.

\section{Conclusions}\label{Sec:Conclusions}

In this manuscript, we generated the ``mixed" quadratic term involving both scalar and vector field in a traditional way, similar to \cite{YM4d}, based on the explicitly Lorentz-breaking coupling of the scalar, vector and spinor fields. This term is naturally finite. Then, it turns out to possess a ``restricted" gauge invariance, that is, it is invariant if only the vector field suffers gauge transformations. However, this situation is common in many theories obtained via the dual embedding procedure (see \textit{e.g.} \cite{dual0,dual}). Also, we succeeded to generate this term through the proper application of the Julia-Toulouse methodology. Finally, we studied the dispersion relations in the electrodynamics involving this term as an additive one.

\section*{Acknowledgements} This work was partially supported by Conselho
Nacional de Desenvolvimento Cient\'{\i}fico e Tecnol\'{o}gico (CNPq).  A. Yu. P. has been supported by the CNPq project No. 303438-2012/6 and C. W.  has been supported by the CNPq project No. 305639-2013/7.

\appendix

\section{The Physical Spectra of the ``Mixed'' Model} \label{Sec:Application}

As an application of the perturbative methods discussed in Sec. \ref{Sec:Perturbative}, the one-loop corrected effective Lagrangian of $A^{\mu}$ and $\phi$ being the sum of the classical Lagrangian of these fields, see (\ref{l}) with the one-loop correction given by (\ref{1loop}), looks like
\bea \label{eq:Leff}
L_{eff}=-\frac{1}{4}F_{\mu\nu}F^{\mu\nu}-\frac{1}{2}\phi(\Box+M^2)\phi+\epsilon^{\alpha\mu\nu}F_{\alpha\mu}v_{\nu}\phi,
\eea
where $v_{\nu}=-eg\frac{m}{8\pi|m|}a_{\nu}$. 

Notice that the above effective lagrangian is not the complete model as there are other terms that can potentially contribute to Eq. (\ref{eq:Leff}). In this manuscript these terms  are neglected since we are only interested in the influence of this ``mixed'' term on the physical spectra. 

{Let us briefly discuss the physical spectra of this ``mixed" model.
This theory is a partial case of the theory considered in \cite{Ferr,dual} arising through a dimensional reduction of the electrodynamics with the Carroll-Field-Jackiw term. Therefore the propagator and, consequently, dispersion relations in our case are similar to the propagator and dispersion relations found in \cite{Ferr,dual} (however, unlike \cite{Ferr}, we have here $M^2\neq 0$, i.e. the scalar field is massive, but, unlike \cite{dual}, we have $m=0$, i.e. there is no Chern-Simons term). So, we can merely quote the results from \cite{dual}, which allows us to write down the propagators in the form 
\bea
\label{props}
<A^{\mu}A^{\nu}>&=&(\Delta_{11})^{\mu\nu}=[(\Box-M^2)M_{\mu\nu}-T_{\mu}T_{\nu}]^{-1}(\Box-M^2)\nonumber\\
<\phi\phi>&=&\Delta_{22}=[(\Box-M^2)M_{\mu\nu}-T_{\mu}T_{\nu}]^{-1}M_{\mu\nu}\nonumber\\
<A^{\mu}\phi>&=&-<\phi A^{\mu}>=\Delta_{12}^{\mu}=-\Delta^{\mu}_{21}=-T_{\nu}[(\Box-M^2)M_{\mu\nu}-T_{\mu}T_{\nu}]^{-1},
\eea
so, the problem is reduced to finding the operator $\Delta^{\mu\nu}=[(\Box-M^2)M_{\mu\nu}-T_{\mu}T_{\nu}]^{-1}$ (with $M_{\mu\nu}=\Box\theta_{\mu\nu}+\frac{\Box}{\xi}\omega_{\mu\nu}$) which we do with use of a special ansatz \cite{dual,JTred}
\bea
\label{ansatz}
\Delta^{\nu\alpha}&=&a_1\theta^{\nu\alpha}+a_2\omega^{\nu\alpha}+a_3S^{\nu\alpha}+a_4\Lambda^{\nu\alpha}+a_5T^{\nu}T^{\alpha}+a_6Q^{\nu\alpha}+a_7Q^{\alpha\nu}+a_8\Sigma^{\nu\alpha}+\nonumber\\&+&a_9\Sigma^{\alpha\nu}+a_{10}\Phi^{\nu\alpha}+a_{11}\Phi^{\alpha\nu},
\eea
where $S_{\mu\nu}=\epsilon_{\mu\lambda\nu}\partial^{\lambda}$, $T_{\nu}=S_{\mu\nu}v^{\mu}$,  $\omega_{\mu\nu}=\frac{\partial_{\mu}\partial_{\nu}}{\Box}$ is a longitudinal projector, $\theta_{\mu\nu}=\eta_{\mu\nu}-\omega_{\mu\nu}$ is a transverse projector, 
$Q_{\mu\nu}=v_{\mu}T_{\nu}$, $\Lambda_{\mu\nu}=v_{\mu}v_{\nu}$, $\Sigma_{\mu\nu}=v_{\mu}\partial_{\nu}$, $\Phi_{\mu\nu}=T_{\mu}\partial_{\nu}$, $\lambda=v^{\mu}\partial_{\mu}$.
These coefficients were found in \cite{dual} for $m\neq 0$ and reduce in our case to
\bea
\label{pmcs}
a_1&=&a_2=\frac{1}{\Box(\Box-M^2)};\nonumber\\
a_3&=&a_4=0;\quad\, 
a_5=\frac{1}{\Box(\Box-M^2){\cal R}};\quad\, a_6=a_7=0;\\
a_8&=&a_9=a_{10}=a_{11}=0.\nonumber
\eea
Here we denoted ${\cal R}=\Box(\Box-M^2)-T^2$. 
Proceeding in a manner similar to \cite{Ferr,dual}, beside of the usual dispersion relations $E^2=\vec{p}^2$ and $E^2=\vec{p}^2+M^2$ we also find $(E^2-\vec{p}^2)(E^2-\vec{p}^2-M^2+v^2)+(\vec{v}\cdot\vec{p}-v_0E)^2=0$. The last relation can be physical only if $v^{\mu}$ is space-like.}

\end{document}